\documentstyle[preprint,aps,floats,epsf]{revtex}

\newcommand{\Journal}[4]{{#1} {\bf #2}, #3 (#4)}

\newcommand{\IJA}{{\em Int. J. Mod. Phys.} A}
\newcommand{\NPB}{{\em Nucl. Phys.} B}
\newcommand{\PLB}{{\em Phys. Lett.}  B}

\newcommand{\PRD}{{\em Phys. Rev.} D}

\newcommand{\shit}[2]{\epsfxsize=#1 \epsfbox[10 30 560 790]{./#2}}

\begin{document}

\preprint{$
\begin{array}{l}
\mbox{November~1997} \\ [0.6cm]
\end{array}
$}

\tightenlines

\title{MSSM RADIATIVE CORRECTIONS TO THE \\ TOP PAIR PRODUCTION
PROCESSES \\ AT HADRON COLLIDERS 
\footnote{Talk presented by D. Wackeroth at
the {\em International Workshop on Quantum Effects in the 
Minimal Supersymmetric Standard Model},
Barcelona, Spain, September 9-13, 1997. To appear in the proceedings,
World Scientific, ed. J. Sol{\`a}.\vspace*{-0.4cm}}\\[0.6cm]}

\author{ W. HOLLIK, W. M. M\"OSLE
\footnote{now at MicroStrategy, Inc.\vspace*{-0.4cm}} }

\address{Institut f\"ur Theoretische Physik, Universit\"at Karlsruhe,\\
D-76128 Karlsruhe, Germany\\E-mail: hollik@itpaxp3.physik.uni-karlsruhe.de}

\author{ C. KAO }

\address{Department of Physics, University of Wisconsin,\\
Madison, Wisconsin 53706, USA\\E-mail: kao@pheno.physics.wisc.edu}

\author{ D. WACKEROTH
\footnote{Address after October 1, 1997: \\
Paul Scherrer Institute, Theory Group F1, CH-5232 Villigen PSI, 
Switzerland\\E-mail: Doreen.Wackeroth@psi.ch} }

\address{Theory Group, Fermi National Accelerator Laboratory,\\
Batavia, Illinois 60510, USA\\E-mail: dow@fnth09.fnal.gov}

\maketitle

\begin{abstract}
We calculate the MSSM ${\cal O}(\alpha)$ corrections to
the main production processes of
strong top pair production at the upgraded Tevatron and at the
LHC. In exploring the potential of future hadron colliders
for performing electroweak precision physics 
we study their effects on the total hadronic cross sections,
the invariant $t\bar t$ mass distributions and
give first results for parity violating asymmetries in the production
of left and right handed top quarks.
\end{abstract}
\section{Introduction}
At future hadron colliders such as the upgraded
Fermilab Tevatron with $\sqrt{S}=2$ TeV 
and the CERN LHC (14 TeV) annual top yields of roughly
$68\cdot 10^3$ (with $\int {\cal L}dt=10 fb^{-1}$) and $10^7$ 
top quark pairs, respectively, are expected \cite{topacc}.
This translates into a measurement of the top quark mass and of the
total $t\bar t$ production rate with a precision of at least
$\delta m_t=2$ GeV and 
$\delta\sigma_{t\overline t}/\sigma_{t\overline t}=6 \%$, respectively.
This envisaged high precision
opens a new rich field of top quark related phenomenology. 
It suggests to explore the potential of performing 
electroweak precision studies
with top quark observables even in an hadronic environment 
in the spirit of the successful LEPI precision physics 
program.
Complementary to direct searches for signals of physics beyond the 
Minimal Standard Model (MSM),
observables being sensitive to the virtual presence of
non-standard particles in strong top pair production processes
might reveal information on the parameter space of the model
under consideration. Especially the strong top-Yukawa coupling
can serve as a probe of the MSM Higgs-sector. Since
the latter is the least experimentally explored sector 
of the MSM, possible extensions are of particular interest.
The consideration of a second Higgs-doublet plays a special
role since it is the minimally required scenario for
mass generation in a supersymmetric extension of the MSM.
Supersymmetry represents an additional symmetry between 
fermions and bosons which implemented in the MSM 
solves such MSM deficiencies like the necessity of fine tuning 
and the non-occurrence of gauge coupling
unification at high energies. 
    
In this contribution we study electroweak-like
quantum effects of the Minimal Supersymmetric Standard Model (MSSM)
in the main top quark production mechanisms 
at the upgraded Tevatron  and at the LHC,
$q \bar q$ annihilation and gluon fusion, respectively.
We give numerical results for the impact of the ${\cal O}(\alpha)$ 
contribution on the
total cross section $\sigma_{t\bar t}$ and on the
invariant mass distribution $d\sigma/d M_{t \bar t}$.
We also present first results for a parity violating asymmetry $A_{LR}$
observed in the strong production of left and
right handed top(antitop) quarks.
For a more detailed discussion 
we refer to \cite{diplpubsm,mssm} and \cite{asym}.
\section{MSSM quantum effects in top pair production
at hadron colliders}
At ${\cal O}(\alpha)$ within the MSSM the $gt\bar t$-vertex
is modified through the virtual presence 
of a heavy and a light neutral scalar, $H^0$ and $h^0$, 
a pseudo scalar $A^0$ and a charged Higgs boson $H^{\pm}$ as well as
charginos $\tilde \chi^{\pm}_i$ and sbottoms $\tilde b_{L,R}$ and neutralinos
$\tilde \chi^0_i$ and stops $\tilde t_{1,2}$. 
For the sake of completeness we also take into account the 
contributions of the electroweak gauge bosons ($W$ and $Z$ bosons).
The ${\cal O}(\alpha)$ contribution can be parametrized in terms of 
form factors revealing the Lorentz-structure 
of the matrix elements to top pair production 
processes as follows:

\underline{$q\bar q$ annihilation:}
\begin{eqnarray}\label{eq:one}
\delta {\cal M}_{q\bar q}^{fin}
& = &\alpha \alpha_s  \frac{i T^c_{ik} T^c_{jl}}{\hat s} \bar{u}_t^j(p_2,s_2)
\Bigl[\gamma_{\mu} \, (F_V +\gamma_5 \, G_A) \Bigr.
\nonumber\\
&+& \Bigl. (p_1-p_2)_{\mu}\, \frac{1}{2 m_t} F_M  \Bigr]
 v_{\bar t}^l(p_1,s_1) 
\, \bar{v}_{\bar q}^k(p_3,s_3)\gamma^{\mu} \, u_{q}^i(p_4,s_4)
\end{eqnarray}

\newpage

\underline{gluon fusion:}\\

\noindent
The ${\cal O}(\alpha)$ contribution to the gluon fusion subprocess
comprises the vertex corrections in the $s$-and $t,u$-production channel
$F_{V,M},G_A$ and $\rho_{i,(V,A)}^{V,(t,u)}$, respectively, the 
off-shell top self energy insertion $\rho_{i,(V,A)}^{\Sigma,(t,u)}$,
the box diagrams $\rho_{i,(V,A)}^{\Box,(t,u)}$ and the 
$s$-channel Higgs-exchange diagrams $\rho_{12}^{\triangleleft}$
\begin{eqnarray}\label{eq:two}
\delta {\cal M}_{gg} &  = &  \alpha \, \alpha_s \, \left\{
\frac{f_{abc} T^c_{jl}}{\hat s} \, 
\left[ (M_2^{V,t}-2 M_3^{V,t}) \, F_V+
(M_2^{A,t}-2 M_3^{A,t}) \, G_A
\right. \right.
\nonumber\\
&+& \left. \left. 
( (\hat t-\hat u) \, M_{12}^{V,t}-4 M_{15}^{V,t}+4 M_{17}^{V,t} )
\, \frac{F_M}{2 m_t^2}\right]
\right.
\nonumber\\
&+ & \left. \sum_{L=V,A} \sum_{i=1,\ldots,7 \atop 11,\ldots,17}
\Bigl( i \,T^a_{jm} T^b_{ml} \, M_i^{L,t} \,\left[\frac{\rho_{i,L}^{V,t}}{\hat t-m_t^2}+\frac{\rho_{i,L}^{\Sigma,t}}{(\hat t-m_t^2)^2}
+\rho_{i,L}^{\Box,t} \right] \Bigr. \right.
\nonumber\\
&+& \left. \Bigl.  i \,T^b_{jm} T^a_{ml} \, M_i^{L,u} \, 
[\hat t\rightarrow \hat u] \Bigr)
+ \sum_{S=H^0,h^0} \frac{(-i \, \delta^{ab}) \, M_{12}^{V,t} \,
\rho_{12}^{\triangleleft}(\hat s,M_S)}
{(\hat s-M_S^2)^2+ (M_S \,\Gamma_S)^2} \right\},
\end{eqnarray}
where $\hat s,\hat t,\hat u$ are Mandelstam variables and
$T^c=\lambda^c/2$ with the Gell-Mann-matrices $\lambda^c$. The $f_{abc}$
are the $SU(3)$ structure constants.
The explicit expressions for the form factors and the standard matrix elements 
$M_i^{(V,A),(t,u)}$ can be found in \cite{diplpubsm,mssm,asym}.
Contracting $\delta M_{q\bar q,gg}$
with the Born-matrix elements yields the partonic 
differential cross sections $d \hat {\sigma}_{q\bar q,gg}/d \hat t$
to ${\cal O}(\alpha \alpha_s^2)$.
The observable hadronic cross section is
obtained by convoluting the partonic cross sections
with parton distribution functions (after performing the $\hat t$ integration) 
\begin{equation}
\sigma_{t\bar t}(S) = \sum_{ij=q\bar q,gg} \int_{\frac{4 m_t^2}{S}}^1 \frac{d\tau}{\tau}
\left(\frac{1}{S} \frac{dL_{ij}}{d\tau} \; \hat s \hat \sigma_{ij}(\hat s,\alpha_s(\mu)) \right)
\end{equation}
with $\tau=x_1 x_2=\hat s/S$ and the parton luminosities
denoted by $L_{ij}$.
In the numerical evaluation
we use the MRSA set of parton distribution functions \cite{mrsa}
with the factorization ($Q$) and renormalization scale ($\mu$) chosen to be
$Q=\mu=m_t$ with $m_t=175$ GeV. 
We also require for the transverse momentum of the top quark that 
$p_t>20,100$ GeV (Tevatron, LHC). 
\subsection{The total cross section $\sigma_{t\overline t}$}\label{subsec:tot}
In order to reveal the numerical impact of the 
MSSM ${\cal O}(\alpha)$ contribution on the observable cross sections
we introduce the relative correction $\Delta$
\begin{equation}
\sigma_{t\bar t}(S) = \sigma_B(S)+ \delta \sigma(S)=\sigma_B (1+\Delta) \;.
\end{equation}
In the numerical discussion \cite{mssm} we study the 
dependence of $\Delta$ on the MSSM
input parameters which we chose to be
\[\tan\beta,M_{A^0},\mu,M_2,m_{\tilde b_L},m_{\tilde t_2},\Phi_{\tilde t}
 \; .\]
\begin{figure}
\begin{center}
\begin{picture}(16,7)
\put(2,4.8){\makebox(0,0){\shit{6.8cm}{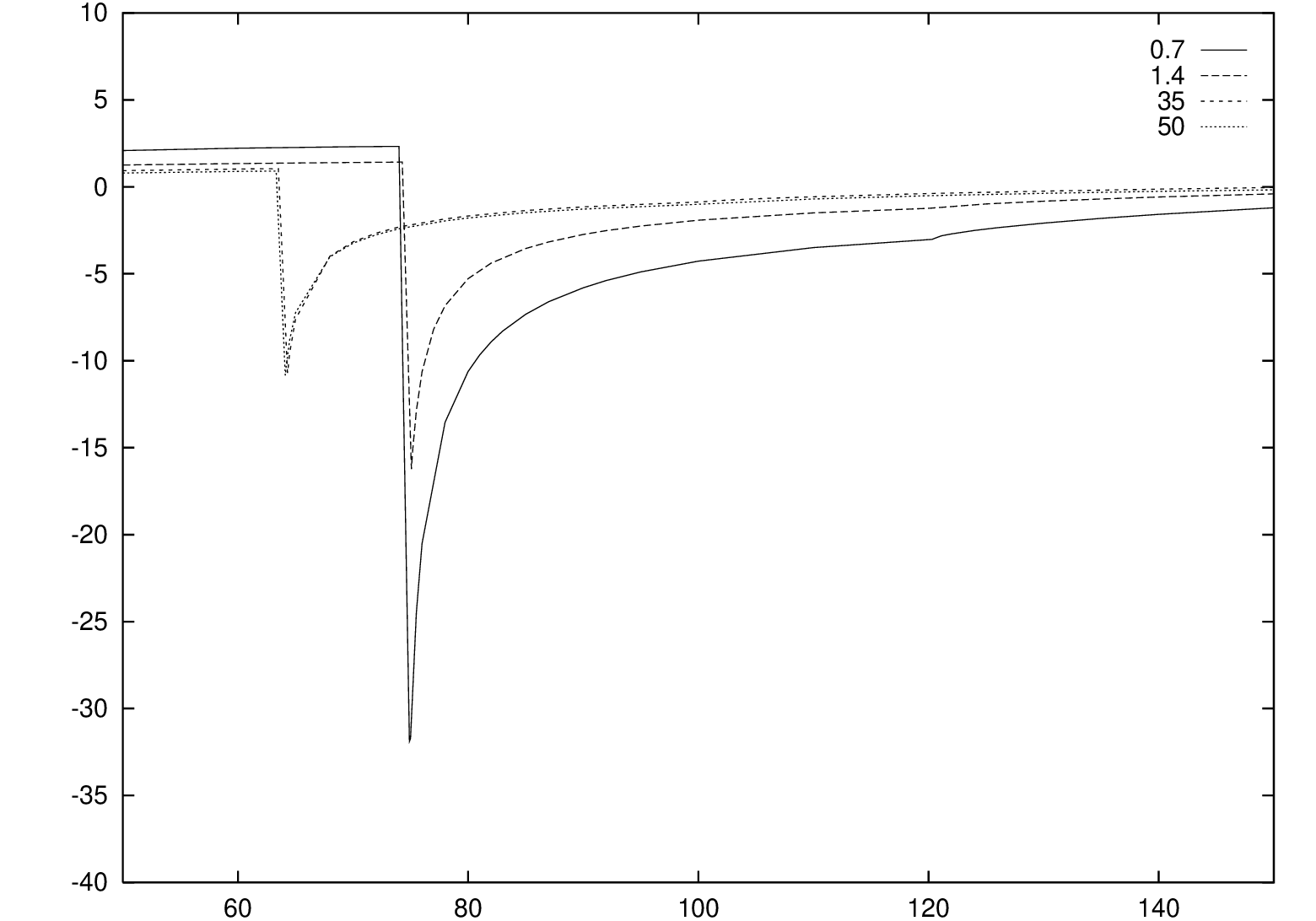}}}
\put(0,7){\makebox(0,0){$\Delta , \%$}}
\put(3.25,0){\makebox(0,0){$m_{\tilde t_2}$ , GeV}}
\put(4,7){\makebox(0,0){Tevatron}}
\put(10.5,4.8){\makebox(0,0){\shit{6.8cm}{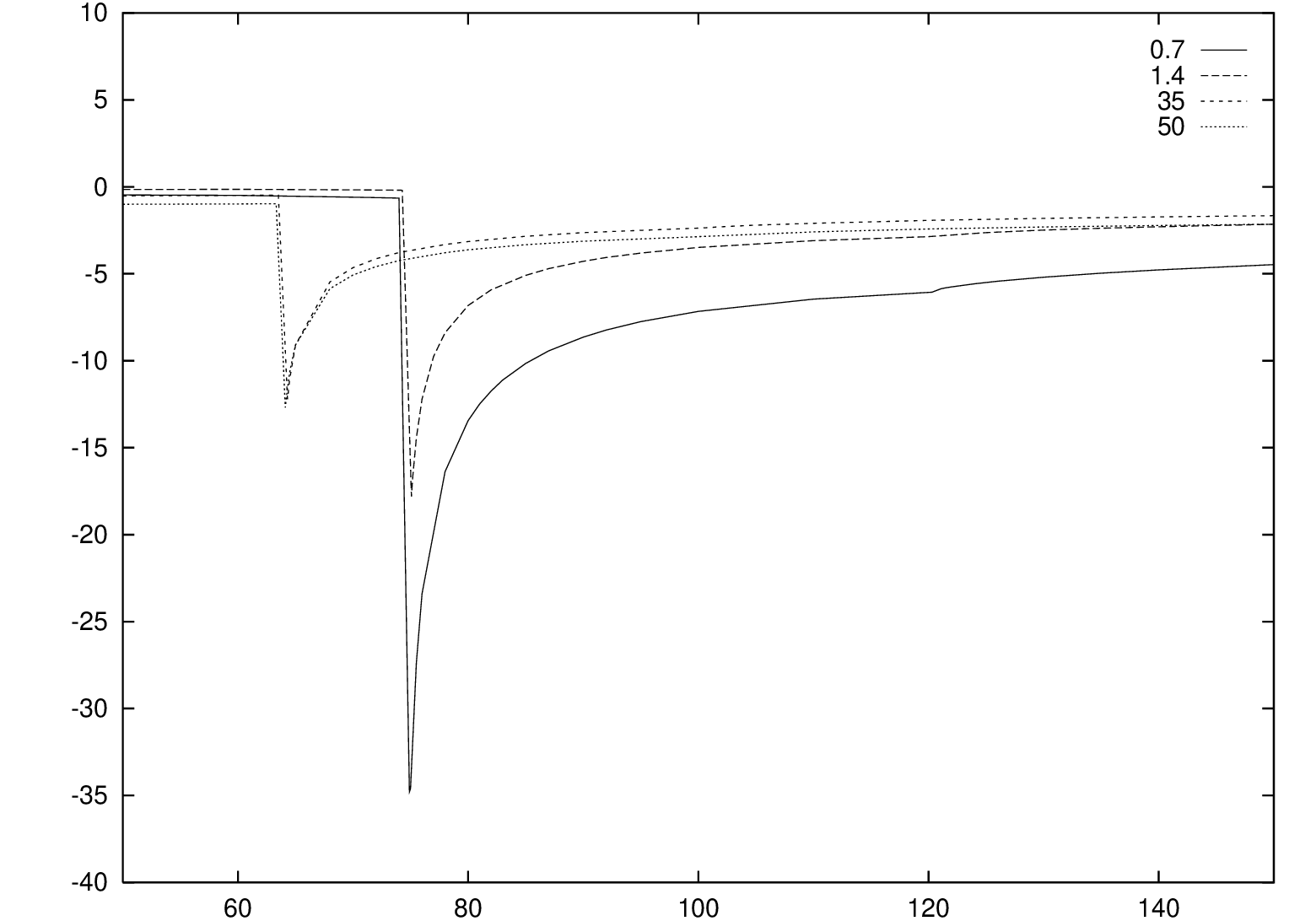}}}
\put(8.5,7){\makebox(0,0){$\Delta , \%$}}
\put(12.5,0){\makebox(0,0){$m_{\tilde t_2}$ , GeV}}
\put(12,7){\makebox(0,0){LHC}}
\end{picture}
\end{center}
\caption{The variation of $\Delta$
with $m_{\tilde t_2}$ and $\tan\beta$ at the Tevatron and at the LHC
when only including the SUSY EW-like corrections
(with $\mu=100$ GeV, $M_2=3 |\mu|$, $m_{\tilde b_L}=150$ GeV and $\Phi_{\tilde t}=\pi/4$).}
\label{fig:one}
\end{figure}\vspace*{0.5cm}
We consider $L,R$-mixing only in the stop-sector ($\Phi_{\tilde t}=0$
no mixing) with $\tilde t_2$ denoting the lighter stop
and we assume $m_{\tilde b_R}=m_{\tilde b_L}$.
The only GUT-constraint we make use of is the relation $M_1=5/3 
\tan\theta_W M_2$.
The most pronounced effect of the MSSM corrections is
illustrated with Fig.~\ref{fig:one}. There the electroweak-like 
SUSY contribution (=MSSM without the Higgs-sector)
reduces the observable cross sections up to
about $35\%$
when approaching the threshold for the decay $t\rightarrow \tilde t_2 
\tilde\chi^0$. For most of the parameter space the 
MSSM ${\cal O}(\alpha)$ contribution
reduces the observable cross sections by several per cent, typically up to
$5\%$ at the Tevatron and up to $10\%$ at the LHC.
\subsection{The invariant mass distribution $\frac{d\sigma}{d M_{t\overline t}}$}\label{subsec:inv}
\begin{figure}
\begin{center}
\begin{picture}(16,7)
\put(2,4.8){\makebox(0,0){\shit{6.8cm}{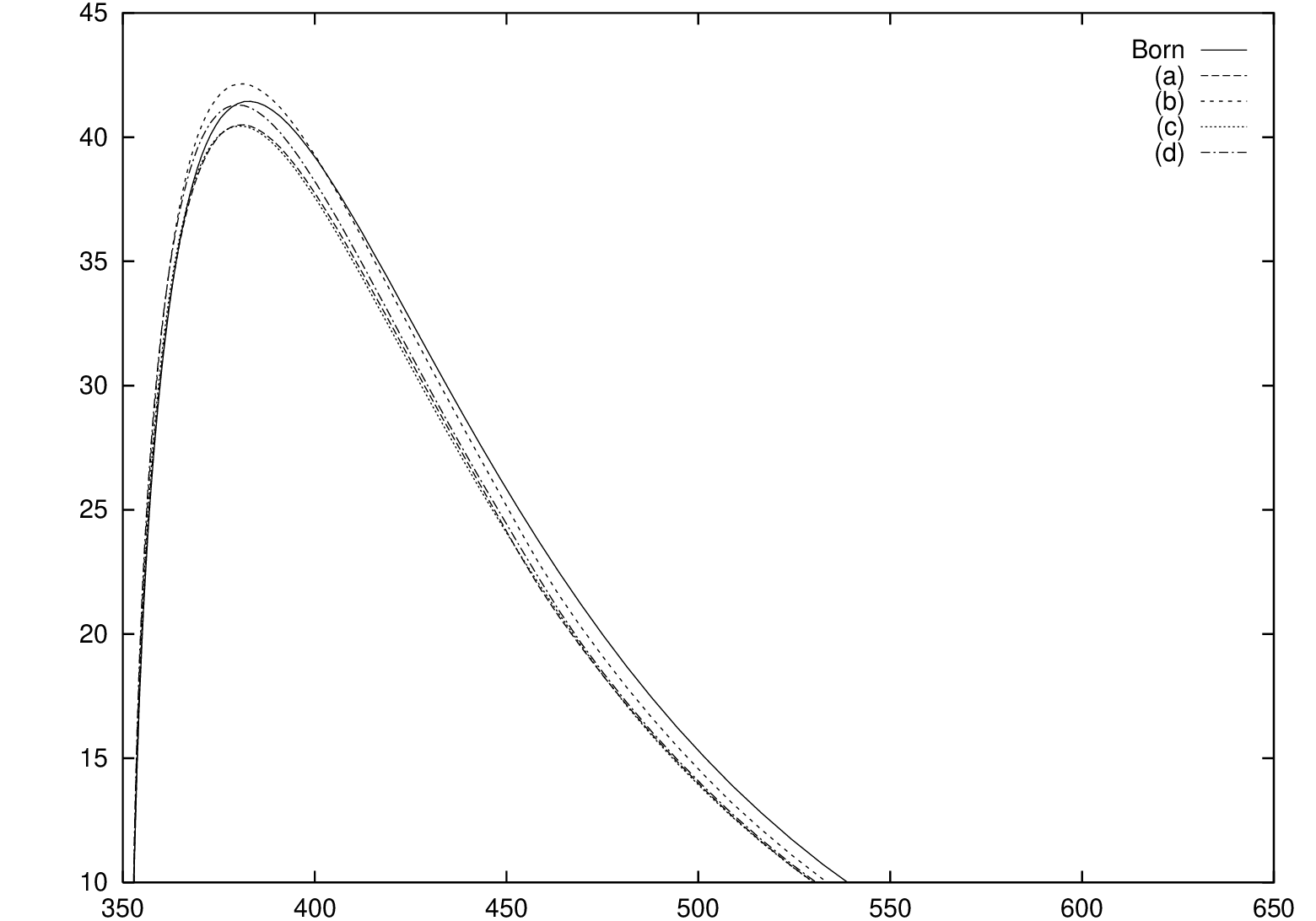}}}
\put(0,7){\makebox(0,0){$\Delta , \%$}}
\put(3.25,0){\makebox(0,0){$M_{t\overline{t}}$ , GeV}}
\put(4,7){\makebox(0,0){Tevatron}}
\put(10.5,4.8){\makebox(0,0){\shit{6.8cm}{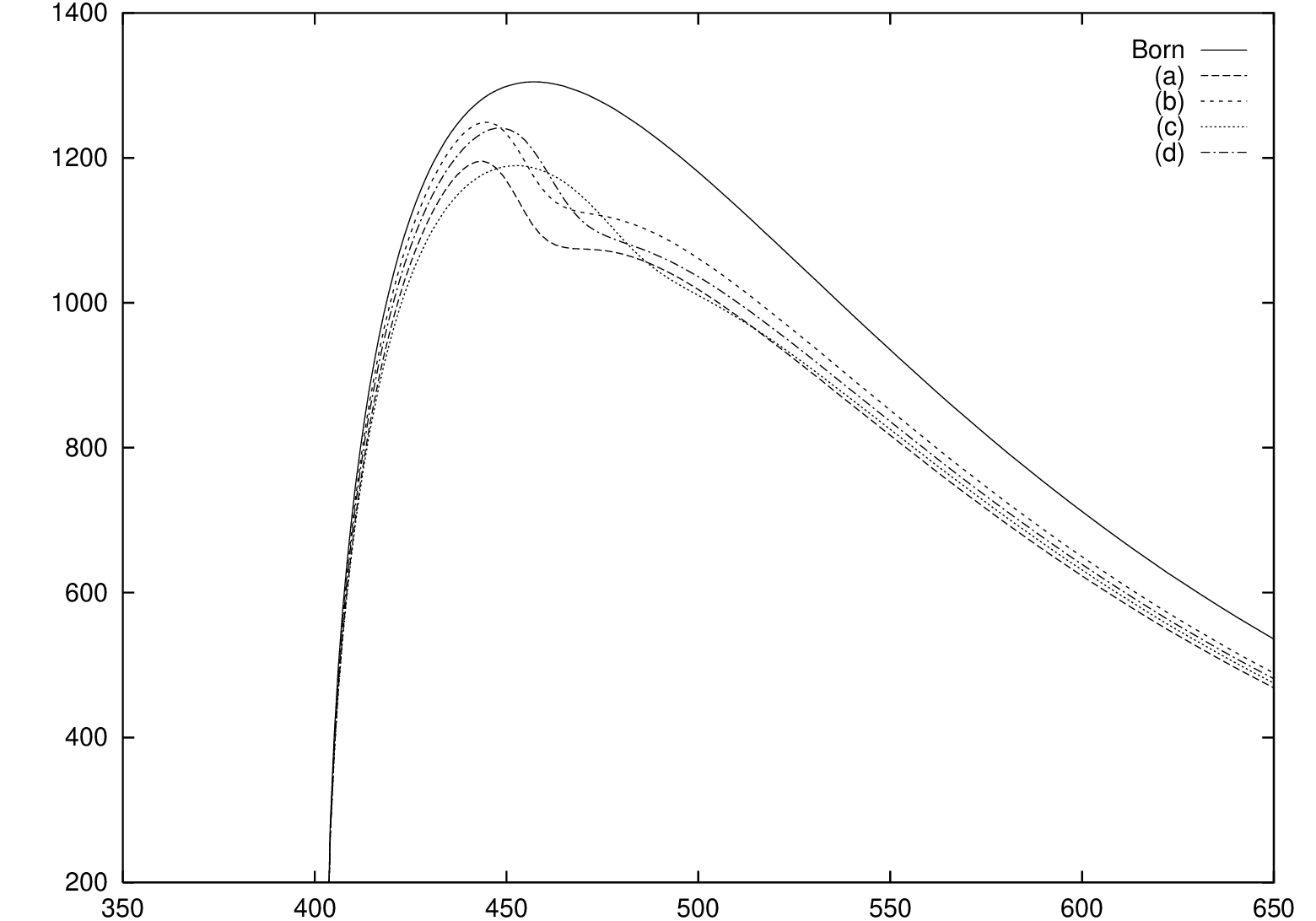}}}
\put(8.5,7){\makebox(0,0){$\Delta , \%$}}
\put(12.5,0){\makebox(0,0){$M_{t\overline{t}}$ , GeV}}
\put(12,7){\makebox(0,0){LHC}}
\end{picture}
\end{center}
\caption{The invariant mass $M_{t \overline t}$
distribution within different MSSM
scenarios (with $\tan\beta=0.7, M_A=450$ GeV, $M_2=3 |\mu|$ and 
with (a) $m_{\tilde t_2}=50$ GeV, $m_{\tilde b_L}=150$ GeV, $\Phi_{\tilde t}=\pi/4$, $\mu=-120$ GeV,
(b) $m_{\tilde t_2}=75$ GeV, 
(c) $m_{\tilde t_2}=75$ GeV, $m_{\tilde b_L}=400$ GeV, $\Phi_{\tilde t}=-\pi/4$, $\mu=150$ GeV
and (d) $m_{\tilde t_2}=75$ GeV, $m_{\tilde b_L}=800$ GeV, $\Phi_{\tilde t}=0$, $\mu=150$ GeV).}
\label{fig:two}
\end{figure}

In Fig.~\ref{fig:two} we show the impact of
the MSSM ${\cal O}(\alpha)$ contribution
on the invariant mass distribution of
the produced top quark pair
\begin{equation}
\frac{d\sigma}{d M_{t\bar t}} = \sum_{ij=q\bar q,gg}
\frac{2}{M_{t\bar t}}\, \hat\sigma_{ij}(\hat s=\tau S) \tau 
\frac{dL_{ij}}{d\tau}
\end{equation}
with $\tau=M_{t \bar t}^2/S$.
There we are especially interested in 
the distortion of the $t \bar t$ invariant mass due to the
$s$-channel Higgs-exchange diagrams in the gluon fusion subprocess
as a very characteristic
signature of the electroweak symmetry breaking sector
possibly observable at $pp$ colliders.
As can be seen in Fig.~\ref{fig:two} at the LHC a significant distortion 
arises for $M_{H^0}>2 m_t$ for a sufficiently small
Higgs-decay width $\Gamma_{H^0}$.
\newpage
\subsection{The differential asymmetry $A_{LR}$}\label{subsec:asym}
\begin{figure}
\begin{center}
\begin{picture}(16,7)
\put(2,4.8){\makebox(0,0){\shit{6.8cm}{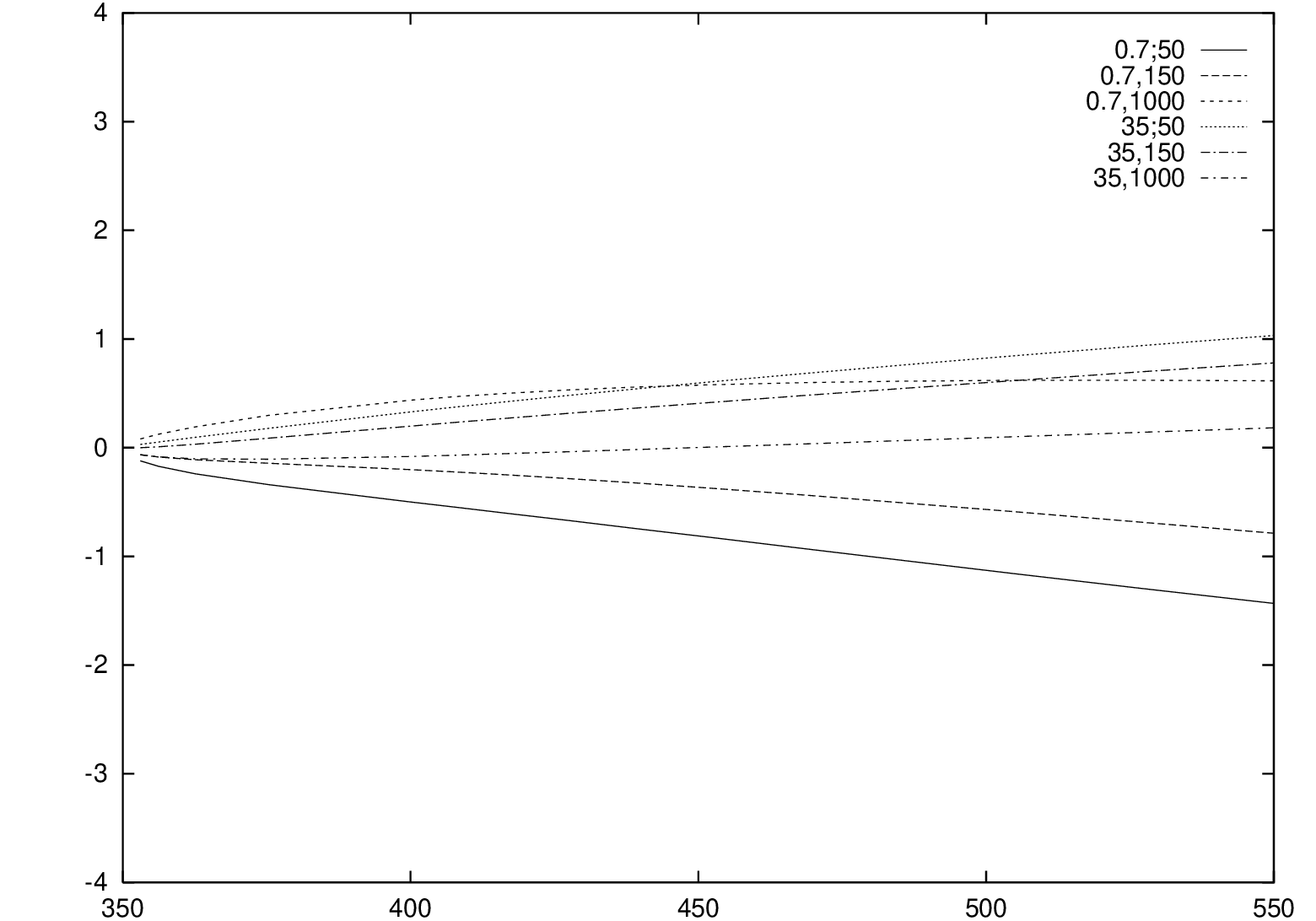}}}
\put(0,7){\makebox(0,0){$A_{LR} , \%$}}
\put(3.25,0){\makebox(0,0){$M_{t\overline{t}}$ , GeV}}
\put(4,7){\makebox(0,0){Tevatron}}
\put(10.5,4.8){\makebox(0,0){\shit{6.8cm}{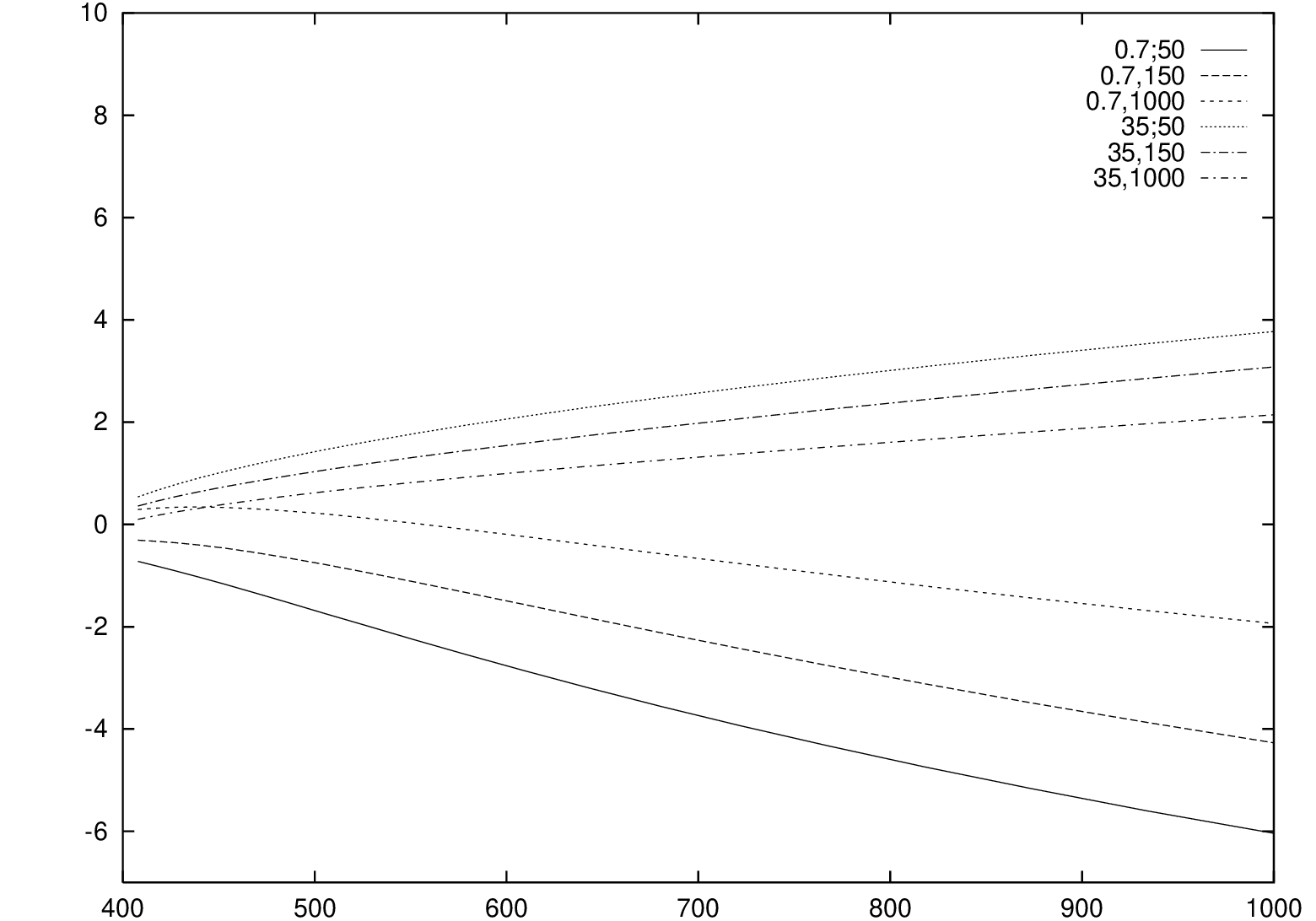}}}
\put(8.5,7){\makebox(0,0){$A_{LR}, \%$}}
\put(12.5,0){\makebox(0,0){$M_{t\overline{t}}$ , GeV}}
\put(12,7){\makebox(0,0){LHC}}
\end{picture}
\end{center}
\caption{$A_{LR}$ as a function of $M_{t \overline t}$ within the MSSM
for different choices for $\tan\beta$ and $M_{A^0}$ (with  $\mu=-120$ GeV,
$M_2=3 |\mu|$, $m_{\tilde t_2}=75$ GeV, $m_{\tilde b_L}=150$ GeV, 
$\Phi_{\tilde t}=\pi/4$).}
\label{fig:three}
\end{figure}

We calculated the helicity amplitudes to the $q \bar q$ annihilation 
and gluon fusion subprocesses to ${\cal O}(\alpha \alpha_s^2)$
by using the helicity projection operators 
\begin{eqnarray}
\bar u_t(p_2) u_t(p_2)& = & (1+2\lambda_t \gamma_5 \not\!{s}_t) 
(\not\!{p}_2+m_t)/2 \nonumber\\
v_{\bar t}(p_1) \bar v_{\bar t}(p_1) & = &
(1+2\lambda_{\bar t} \gamma_5 \not\!{s}_{\bar t})(\not\!{p}_1-m_t)/2
\end{eqnarray}
when contracting $\delta {\cal M}_{q\bar q,gg}$ of 
Eq.~\ref{eq:one},\ref{eq:two} with the Born-matrix elements.
$\lambda_{t,\bar t}=\pm 1$ denotes the top/antitop helicity. 
Here we only present results for the case when 
the particle's spin is decomposed along its direction of motion.
In \cite{asym} we will also discuss the effects 
of different choices for the spin four-vector $s_{t,\bar t}$
as inspired by the discussion in 
\cite{yael}.
We define a differential left-right asymmetry \cite{chung}
\begin{equation}
A_{LR} = \frac{d\sigma_{RL}/d M_{t\bar t}-d\sigma_{LR}/M_{t\bar t}}
{d\sigma_{RL}/d M_{t\bar t}+d\sigma_{LR}/M_{t\bar t}} \; ,
\end{equation}
where $\sigma_{RL}$ denotes the cross section for the 
production of a top quark with $\lambda_t=+1$ ($t_R$) together with
an antitop quark with $\lambda_{\bar t}=-1$ ($\bar t_L$).
$A_{LR}$ is zero at the Born-level because QCD conserves parity
and small ($<1\%$) when induced 
through parity violating interactions within the MSM.
Therefore, it is a promising candidate for observing
effects of non-minimal SM loop-induced parity violation.
In Fig.~\ref{fig:three} we show $A_{LR}$ obtained within the
MSSM for a characteristic choice of the input parameters.
\newpage
\section{Conclusions}
We studied the effects of MSSM ${\cal O}(\alpha)$ corrections
in top pair production 
at the upgraded Tevatron and at the LHC.
The observable hadronic cross sections
are typically reduced by several percent ($\stackrel{<}{\sim}10 \%$)
but in exceptional regions of the parameter space,
that is when $m_t\approx m_{\tilde t_2}+m_{\tilde\chi^0}$,
the radiative corrections are considerably enhanced,
comparable in size to QCD effects. As a characteristic signature
of the MSSM Higgs-sector a distortion of the 
$t\bar t$ invariant mass distribution
can be observed originating from a resonance in the gluon fusion
subprocess.
First results for a differential left-right asymmetry
loop-induced by parity violating electroweak-like interactions within the MSSM
are promising enough to motivate a more detailed study of the
potential to use the top spin information 
for the detection/exclusion of specific non-standard physics scenarios. 
We conclude that 
provided the intrinsic QCD uncertainties can be considerably
reduced there is potential for electroweak precision studies
in strong processes at future hadron colliders.
\acknowledgements
D. W.~would like to thank S. Parke and Y. Shadmi for
helpful discussions.

The Fermi National Accelerator Laboratory is operated by 
the Universities Research Association, Inc., under contract
DE-AC02-76CHO3000 with the United States Department of Energy.


\begin{references}
%
\bibitem{topacc}
Report of the {\em tev2000} Study Group, eds. D. Amedei and R. Brock
(FERMILAB-Pub-96/082, April 1996).
%
\bibitem{diplpubsm}
W. Beenakker, A. Denner, W. Hollik, R. Mertig, T. Sack, and D. Wackeroth,
\Journal{\NPB}{411}{343}{1994}.
%
\bibitem{mssm}
W. Hollik, W.M. M\"osle, and D. Wackeroth,
hep-ph/9706218, to be published in {\em Nucl. Phys.} B.
%
\bibitem{asym}
C. Kao and D. Wackeroth, in preparation.
%
\bibitem{mrsa}
A.D. Martin, R.G. Roberts, and W.J. Stirling,\Journal{\PRD}{50}{6734}{1994}.
%
\bibitem{yael}
G. Mahlon and S. Parke, hep-ph/9706304 and references therein.
%
\bibitem{chung}
C. Kao, \Journal{\PLB}{348}{155}{1995}; \\
C. Kao, G.A. Ladinsky, and C.P. Yuan, \Journal{\IJA}{12}{1341}{1997};\\
C.S. Li, R.J. Oakes, J.M. Yang, and C.P. Yuan, \Journal{\PLB}{398}{298}{1997}.
%
\end{references}
\end{document}